# Physics of Silicene Stripes

A. Kara, C. Léandri, M. E. Dávila, P. de Padova, B. Ealet, H. Oughaddou,

B. Aufray, and G. Le Lay

**Abstract** Silicene, a monolayer of silicon atoms tightly packed into a two-dimensional honeycomb lattice, is the challenging hypothetical reflection in the silicon realm of graphene, a one-atom thick graphite sheet, presently the hottest material in condensed matter physics. If existing, it would also reveal a cornucopia of new physics and potential applications. Here, we reveal the epitaxial growth of *silicene stripes* self-aligned in a massively parallel array on the anisotropic silver (110) surface. This crucial step in the silicene « gold rush » could give a new kick to silicon on the electronics road-map and opens the most promising route towards wide-ranging applications. A hint of superconductivity in these *silicene stripes* poses intriguing questions related to the delicate interplay between paired correlated fermions, massless Dirac fermions and bosonic quasi-particules in low dimensions.

**Keywords** Silicene, graphene, stripes, superconductivity

C. Leandri . B. Ealet . B. Aufray . G. Le Lay (✉)
CINaM-CNRS, Campus de Luminy, Case 913, F-13288 Marseille Cedex 09, France
e-mail: lelay@cinam.univ-mrs.fr
H. Oughaddou, Departement de Physique, University de Cergy-Pontoise, France.
**A. Kara,** Department of Physics, University of Cental Florida ,
Orlando FL 32816, USA
**M.E. Dávila,** Instituto de Ciencia de Materiales de Madrid, CSIC, Cantoblanco 28049 Madrid, Spain
**P. de Padova,** ISM-CNR, via del Fosso del Cavaliere, 00133 Roma, Italy

## 1 Introduction

Graphene, the two-dimensional (2D) single layer building sheet of graphite, is the hottest new material in physics and nanotechnology; it has striking exotic properties, which open potentially novel routes for many applications[1-3]. However, « graphenium » microprocessors are unlikely to appear for the next 20 years[4], since replacement of silicon electronics is a tough hurdle. Instead, silicon-based nanotechnology, which is compatible with conventional silicon microtechnology, is highly promising. Hence, there has been considerable interest in the fabrication, characterization and properties of silicon nanowires (SiNWs) and nanodots[5]. Silicene[6], the counterpart of graphene in the silicon nano-world, has attracted strong theoretical attention since several years[7]. Typically, like graphene, its charge carriers would be massless relativistic Dirac fermions. Yet, when such sheets of silicon are formed, the energy extra cost to produce curved structures is very low. Hence, while the fabrication of silicon nanotubes (SiNTs), including most presumably single-walled ones has been reported[8,9], the synthesis of silicene remained up to now a virtual quest. Recently, we have proven controlled epitaxial growth of *silicene stripes* (or silicon nanoribbons: SiNRs), possibly a new paradigmatic figure in the silicon realm, through detailed synergetic results of Scanning Tunneling Microscopy/Spectroscopy (STM/STS) experiments and advanced Density Functional Theory calculations[10]. This discovery stems from the search for the atomic structure of the novel one-dimensional (1D) silicon nanostructures (Fig. 1), found recently in Marseille upon controlled growth of silicon on silver (110) surfaces under ultra-high vacuum conditions[11]. They are one-atom thick SiNRs, i.e., true *silicene stripes*[10].

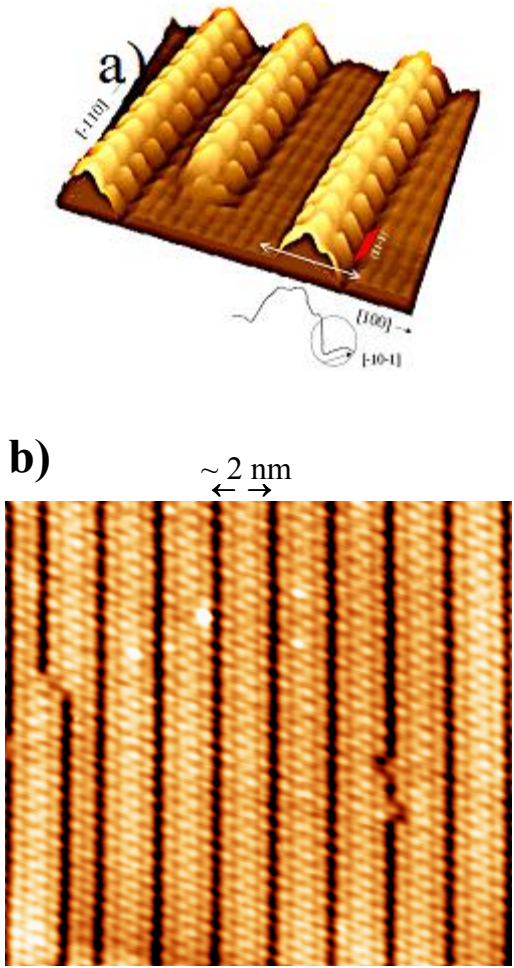

Fig. 1: a) STM image of straight, parallel *silicene stripes* grown at room temperature on a Ag(110) surface (6.2 x 6.2 nm$^2$, I = 1.13 nA, V = -1.7 V, filled states, 3D view). The line profile indicates an asymmetric cross-sectional shape with a clear dip on the right side, creating a (11-1) nanofacet shown in red, b) STM picture of the 4x5 grating with a pitch of just ~ 2 nm (22 x 20 nm$^2$ ; 190 pA ; - 3.3 V).

## 2 Experimental Results

We have studied in details their physical and chemical properties, which are exceptionally striking. First, we mention the straightness and massive parallel alignment along the [-110] channels of the bare surface, giving highly perfect identical (but chiral) quantum-objects (just differing in their lengths) in the macroscopic ensemble over the whole substrate surface. Second, we underline peculiar structural features: (i) the exact "by two" periodicidy of the observed protusions (2 $a_{Ag[1-10]}$ nearest neigbourg distances, that is, 0.578 nm), but with a glide by one Ag-Ag distance between either sides; (ii) the same common height of ~ 0.2 nm related to a slightly arched cross-sectional shape (iii) a « magic » width of ~ 1.7 nm (i. e., about 4 $a_{Ag[001]}$); (iv) an electronic signature reflected in the STM images showing a key structure formed by a square and a diamond side-by-side (Fig. 2); (v) the symmetry breaking across the ribbon with a pronounced dip on one side only ; (vi) the spontaneous separation of the two resulting chiral species into large left-handed and right-handed magnetic-like domains.

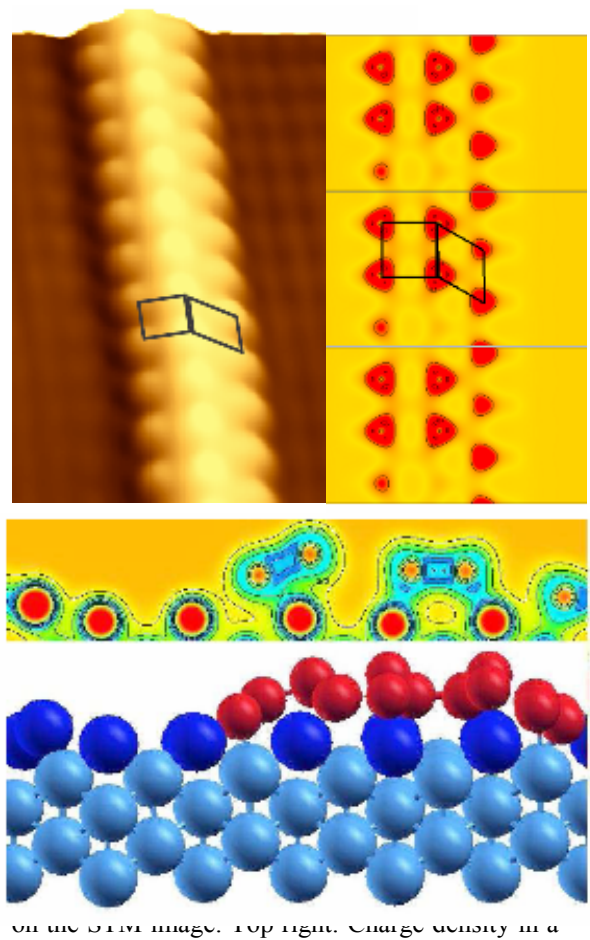

on the STM image. Top right: Charge density in a horizontal plane containing the topmost Si atoms. Middle: Charge density in a vertical plane containing some top most Si atoms. Bottom: Side view showing the curved shape of the stripes (Si in red).

Third, we stress the metallic character and unprecedented spectral signatures: quantized states in the sp region of the valence band below the Fermi level with a strict 1D dispersion along the length, and the narrowest Si 2p core-level lines ever reported in the solid state[11,12] (Fig. 3).

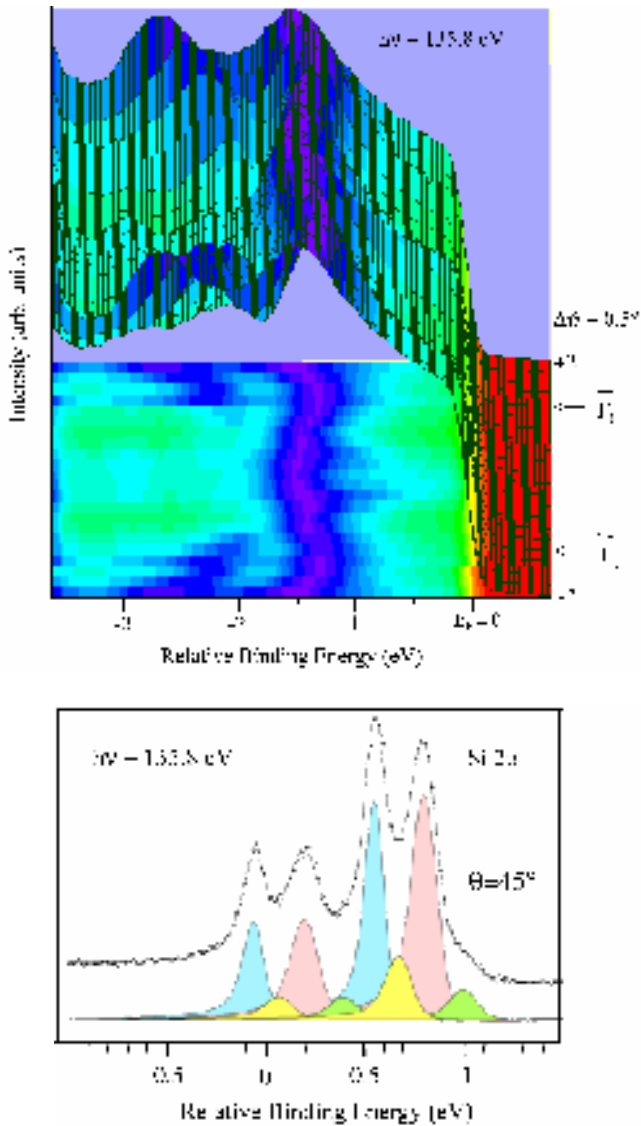

Fig 3: Angle-resolved valence band spectra bellow $E_F$ and 1D dispersion of the quantized states along the stripes (top) and Si 2p core-levels (bottom).

Fourth, we emphasize the amazing *burning match* mechanism of the oxidation process of these stripes, which starts at their extremities[13] (unlike the graphene counterpart which shows strong edge reactivity) making these *silicene stripes* substantially more stable chemically than the graphene based ones (Fig. 4).

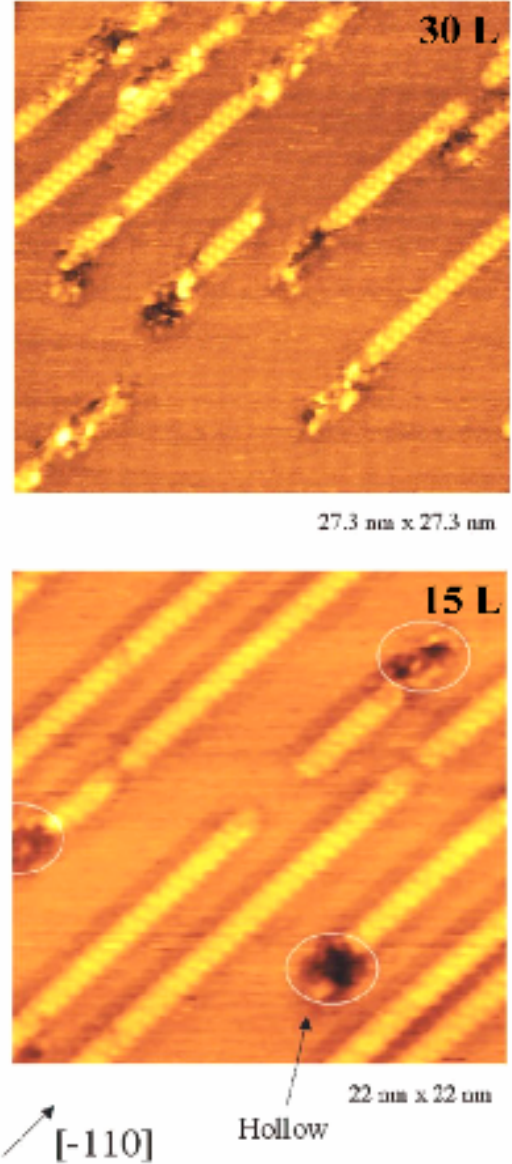

Fig. 4: Filled-states STM images at two different oxygen doses.

The behavior is like a propagating flame front, behind which a ~ 350 meV gap opens on the oxidized part. Close to the moving internal junction thus formed several oxidation states are present, resembling those observed upon the formation of the $SiO_2$/Si(111) interface[14]. From this, we can infer that switching these metallic stripes to n or p-type semiconductors by doping should not be difficult. Finally, we stress that at slightly higher temperatures these unique 1D *silicene stripes* self-assemble by lateral compaction to form a grating at the molecular scale covering the whole substrate surface with a pitch of just ~ 2 nm[15]. To summarize at this point, we have demonstrated the experimental synthesis of silicene, in the form of massively parallel metallic silicon nanoribbons, i.e., *silicene stripes*[10].

The fundamental and strange relativistic effects that are searched for in graphene could be also be studied in silicene[4,5,16]. Especially interesting is the fact that the SiNRs on Ag(110) are intrinsically chiral and that they phase separate into large magnetic-like domains[12]. Could this reflect a symmetry breaking in the quasiparticles in the one-dimensional form of silicene, is an intriguing question, which might be linked to the puzzle of « chiral symmetry breaking » in fundamental physics.

Since to make up a circuit with graphene sheets one envisions to slice them wider or narrower and in different patterns depending on whether a wire, a ribbon or some other component are needed[17], the spontaneous formation *of silicene stripes*, which can further develop into wider, longer and slightly thicker, still massively parallel, SiNRs is rather promising, not to mention the formation of a dense array on the whole surface[11,14]. In this respect, this dense array of massively parallel stripes (giving a 5x4 diffraction pattern) corresponds to an architecture of ~ $5 \times 10^6$ channels/cm, comparing favorably with the active massively parallel architecture of silver adsorbed silicon lines (in a sense the inverse system) on silicon carbide surfaces[18]. With reference to graphene, the epitaxial growth process we have initiated is not so far from the method developed by C. Berger *et al.* for mass-production[19]. Indeed, applications may require further deposition of an insulating support on top and chemical removal of the primary metallic substrate. Very encouraging for practical purposes is the high stability towards oxidation of the SiNTs reported by De Crescenzi *et al.*[8]. In any case, passivation by hydrogen, which should also open a band gap, should be straightforward. All "graphene dreams" that are pursued can be readily transposed to silicene[4,16,17,20]. Hence, improved superconducting (note that the bare SiNRs are metallic, a key point to underline, since metallic silicon, either obtained at high pressure[21] or upon extreme doping with boron[22], is a superconductor; moreover, silver itself, their template, is known to boost superconductivity through an "inverse" proximity effect[23]) and spin-valve transistors could be built, not to mention silicene ballistic transistors. Efficiently, silicene stripe arrays could be prepared in large scale on thin silver (110) epitaxial films grown, e.g., on GaAs(100) substrates[24]. Tantalizing is the possibility that silicene could help prolong the life of Moore's law! Transposing from SiNWs, it could be possible to build ultra-sensitive chemical sensors with *silicene stripes*, e.g., for "Electronic noses" applications[25], bioanalytical applications in aqueous solutions[26], biological[27] and cancer markers[28] detection. We stress that SiNRs have been already used for the formation of well-defined 1D organic nanostructures by selective chemisorption, thus opening the way towards useful functionnalization; as for oxygen, the chemisorption of such organic molecules opens up a 300 meV band gap[29]. We can also anticipate a strong development for solar cells[30], since, already, SiNWs show great potential for this most promising carbon-free power technology[31]. Finally, we emphasize hydrogen storage, which has been an active but controversial subject for carbon nanotubes (CNTs), since most of the efforts have failed to reach the gravimetric density (6.5 wt %) proposed by the U.S. Department of Energy for the hydrogen plan of fuel cell vehicles. It has been suggested that graphene is capable of absorbing a large amont of hydrogen[32], but since it was shown recently that SiNT arrays exhibit much stronger attraction to hydrogen both inside and outside, compared to isodiameter CNTs[33], we can expect that silicene will be far more efficient than graphene.

## 3 Conclusions

In the end, returning to basic physics, we stress the pronounced 1D character of these metallic stripes[11,12]. Consequently, one may conjecture strongly interacting electrons described within the framework of Luttinger liquid, rather than by a Fermi liquid[34]. This may imply a delicate competition between bosonic quasi particles, holons and spinons and the Dirac fermions expected for 2D silicene. Besides, since simple hexagonal metallic silicon is a superconductor at about 8 K, it is not unlikely that *silicene stripes*, i.e., ribbons from a single sheet of such a phase, would be also superconducting, possibly with a Tc increase, as mentioned above. Interestingly, the related silver-germanium couple exhibits striking superconducting interface effects[35,36]. The interplay between a superconducting condensate, Dirac fermions in two dimensions and bosonic quasi-particles in one dimension rises, *per se*, highly intriguing novel questions.

**Acknowledgements This** work was supported by the Ministerio de Educación y Ciencia (Spain), through Research Project MAT2006-13796. The work of AK was supported partially by a UCF start-up fund.